%% file: main.tex
\documentclass{egpubl}
\usepackage{expressive-wiced2025}

\Expressive        % for SBIM, CAe, NPAR
\usepackage[T1]{fontenc}
\usepackage{dfadobe}  

\usepackage{cite}  % comment out for biblatex with backend=biber
\BibtexOrBiblatex
\electronicVersion
\PrintedOrElectronic
\ifpdf \usepackage[pdftex]{graphicx} \pdfcompresslevel=9
\else \usepackage[dvips]{graphicx} \fi
\usepackage{amsfonts}
\usepackage{amssymb}
\usepackage{todonotes}
\usepackage{amsmath}
\usepackage{egweblnk} 
\input{macros}

\usepackage{wrapfig}
\usepackage{algorithm}
\usepackage{algpseudocode}

\title[View-Dependent Deformation Fields for 2D Editing of 3D Models]%
      {View-Dependent Deformation Fields for 2D Editing of 3D Models}

\author[M. El Mqirmi \& N. Aigerman]{
\parbox{\textwidth}{\centering
Martin El Mqirmi \orcid{0009-0005-3285-2085}  and
        Noam Aigerman\orcid{0000-0002-9116-4662} 
        }
        \\
{\parbox{\textwidth}{\centering  Université de Montréal, Canada
       }
}
}
\begin{document}
% make equations vertical space more compact
\setlength{\abovedisplayskip}{10pt}
\setlength{\belowdisplayskip}{10pt}

% uncomment for using teaser
\teaser{
 \includegraphics[width=0.9\linewidth]{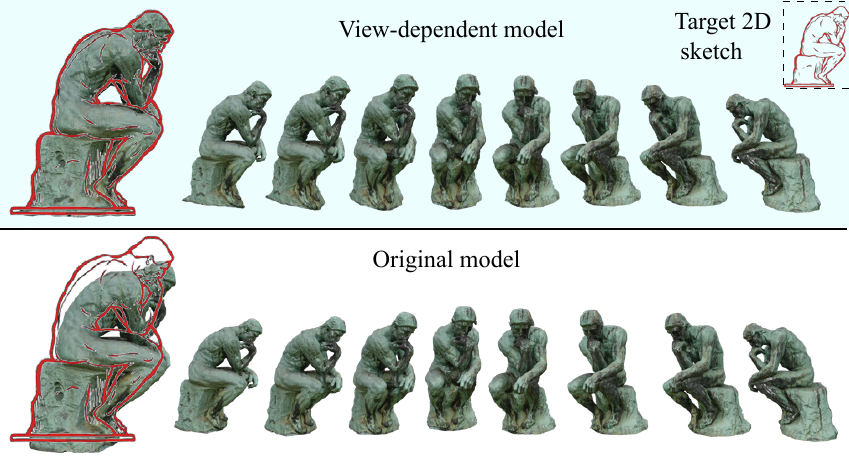}
 \centering
  \caption{\textbf{Thinking from different perspectives: deforming a 3D model of Rodin's sculpture, The Thinker, to match a given 2D sketch (top right, in red), while preserving its appearance from different viewpoints.} The initial model does not align with the 2D sketch (bottom), however our method enables a novice user to perform simple manual 2D warping of the model in 2D, fitting it  perfectly to the sketch (top).  While this naïve 2D warping would have created artifacts in the 3D model when viewed from other directions, this deformation is view-dependent, and its effect vanishes as the camera rotates around the object, making it appear undeformed when viewed from other views.}
\label{fig:teaser}
}

\maketitle

%---------------------------- Abstact ----------------------------------
\input{sections/0_abstract}

%-------------------------- Introduction -------------------------------
\input{sections/1_introduction}

%-------------------------- Related works -------------------------------
\input{sections/2_related_works}

%-------------------------- Method -------------------------------
\input{sections/3_method}

%-------------------------- Experiments -------------------------------
\input{sections/4_experiments}

%-------------------------- Conclusion -------------------------------
\input{sections/5_conclusion}

%------------------------- Bibliography ------------------------------
\bibliographystyle{eg-alpha-doi}  
\bibliography{main}    
% \newpage
\appendix
\input{sections/6_gui}
\input{sections/7_pseudo_code}

\end{document}

%% file: macros.tex
\newcommand{\martin}[1]{#1}

\newcommand{\cezanne}{Cézanne}

\makeatletter
\usepackage{etoolbox}

\newcommand{\myenum}{%
  \ifcsdef{c@myenumcounter}{}{%
    \newcounter{myenumcounter}%
  }%
  \stepcounter{myenumcounter}%
  \textbf{(\alph{myenumcounter})}\ % Print letter with parentheses
}
\makeatother

%% file: sections/0_abstract.tex
\begin{abstract}
   We propose a method for authoring non-realistic 3D objects (represented as either 3D Gaussian Splats or meshes), that comply with 2D edits from specific viewpoints. Namely, given a 3D object, a user  chooses different viewpoints and interactively deforms the object in the 2D image plane of each view. The method then produces a ``deformation field'' - an  interpolation between those 2D deformations in a smooth manner as the viewpoint changes. Our core observation is that the 2D deformations do not need to be tied to an underlying object, nor share the same deformation space. We use this observation to devise a method for authoring view-dependent deformations, holding several technical contributions: first, a novel way to  compositionality-blend between the 2D deformations after lifting them to 3D - this enables the user to ``stack'' the deformations similarly to layers in an editing software, each deformation operating on the results of the previous; second, a novel method to apply the 3D deformation to 3D Gaussian Splats; third, an approach to author the 2D deformations, by deforming a 2D mesh encapsulating a rendered image of the object. We show the versatility and efficacy of our method by adding cartoonish effects to objects, providing means to modify human characters, fitting 3D models to given 2D sketches and caricatures, resolving occlusions, and recreating classic non-realistic paintings as 3D models.
   
%-------------------------------------------------------------------------
%  ACM CCS 1998
%  (see https://www.acm.org/publications/computing-classification-system/1998)
% \begin{classification} % according to https://www.acm.org/publications/computing-classification-system/1998
% \CCScat{Computer Graphics}{I.3.3}{Picture/Image Generation}{Line and curve generation}
% \end{classification}
%-------------------------------------------------------------------------
%  ACM CCS 2012
   % (see https://www.acm.org/publications/class-2012)
%The tool at \url{http://dl.acm.org/ccs.cfm} can be used to generate
% CCS codes.
%Example:

\ccsdesc[500]{Computing methodologies~Shape modeling}

\printccsdesc   
\end{abstract}

%% file: sections/1_introduction.tex
\section{Introduction}

\begin{figure}
    \centering
    \includegraphics[width=\linewidth]{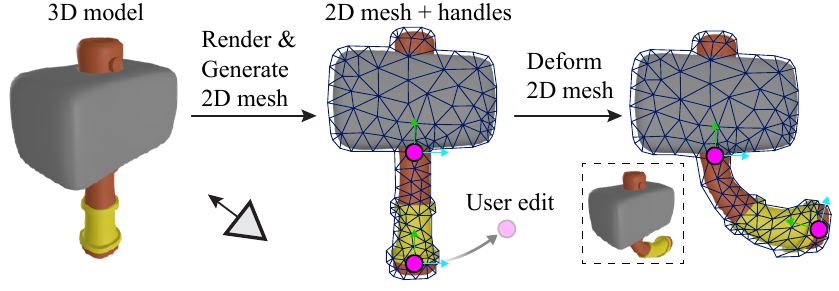}
    \caption{\textbf{Creating a single ``2D'' deformation of a 3D object.} A 3D model (left) is rendered from a chosen view point $v_i$ into a 2D image (middle), which is meshed. The user selects deformation handles (magenta circles) and drags them (``user edit''), creating a deformation of the 2D mesh (right), which defines a 2D deformation $\phi_i$ and is lifted to a 3D deformation $\Phi_i$ (dashed square). }
    \label{fig:2D_deform}
\end{figure}

This paper proposes a technique for 2D-based deformation of 3D models (either 3D Gaussian Splats~\cite{3dgs} or meshes) in a view-dependent manner~\cite{vdg},  so that the resulting 3D models can adapt to desired artistic modifications from specific 2D views, while still preserving fidelity from all possible view directions. This in turn enables both authoring 3D models that cannot be realized by a static 3D object, as well as providing the means to users with no experience with 3D modeling, to perform 2D edits in a 3D-compatible way. 

One immediate motivation for our work stems from modern visual media such as illustrations, caricatures and cartoons, which often exhibit objects that cannot be realized faithfully in 3D, and often change proportions and features when viewed from different positions~\cite{peanuts}. However, we also aim to provide a general technique that is applicable to other media such as, e.g., fine art, considering that view-dependent  distortion of objects is deeply rooted in human aesthetic.  Indeed, throughout most of the history of human civilization, imagery produced by humans has not been faithful to the exact proportions and pose of the underlying object they aimed to represent. Over centuries, what probably originated from a limitation in humans' capacity to accurately and consistently capture 3D geometry from various viewpoints, has become an inherent part of many artistic styles, which in turn would lose their essence if 3D realism were to be enforced.
Earlier examples of blatant non-realistic styles include examples such as Hieroglyphs, and christian scenes from the middle ages.  Artists like Van Gogh (Figure~\ref{fig:van_gogh_chair}) and \cezanne{}  (Figure~\ref{fig:still_life}) have produced works that exhibit more subtle violation of perspective rules,  inspiring later perspective-challenging artistic movements such as Cubism.

Unfortunately, the advent of computer graphics (CG) and 3D digital animation has made these expressive approaches far less prominent. Indeed, the novel tools introduced for authoring 3D CG content (e.g., \textit{Maya} and \textit{Blender}) render actual digital 3D models, and as a result, straightforward use of these tools directly goes against producing any view-dependent inconsistencies, thereby preventing the aesthetic of the traditional 2D approach.
This lacuna is often experienced, for example, by 3D animation studios, especially when they aim to revive a classic 2D-based work and bring it into the 3D CG realm. This leads these studios to develop ad-hoc techniques for adapting 3D content in order to make it appear true to the original 2D artwork, at the price of designing restricted, task-specific frameworks, or otherwise corrupting the 3D content, such that it appears malformed in any other view direction than the one intended for by the 3D artist. An example of this issue is mentioned by the creators of the Peanuts movie in a video interview~\cite{peanuts}.

\begin{figure}
    \centering
    \includegraphics[width=\linewidth]{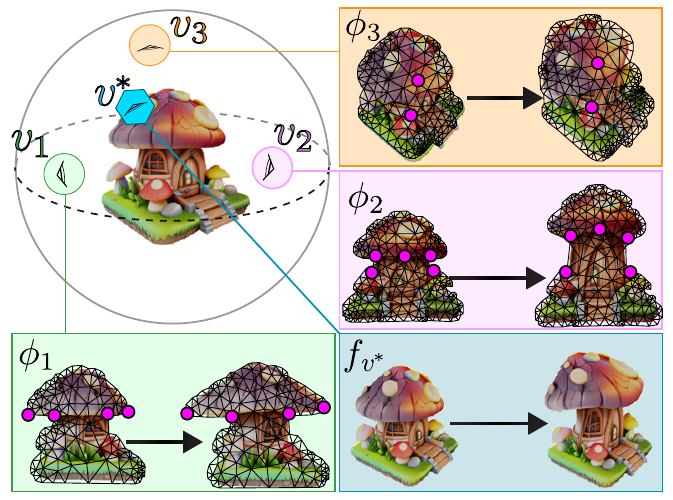}
    \caption{\textbf{Creating a view-dependent deformation \textit{field} from several 2D deformations.}  A 2D mesh of the object is created for each of the keypoint views $v_1,v_2,v_3$. The meshes are deformed by the user to define three 2D deformations, $\phi_1,\phi_2,\phi_3$. When viewed from another view, $v^*$, the three deformations are lifted to 3D and interpolated, using the compositional interpolation formula, Equation~\eqref{eq:interpolation_point}, to yield the 3D deformation $f_{v^*}$. }
    \label{fig:interpolation}
\end{figure}

To tackle this issue, we draw inspiration from the recent emergence of \textit{view-dependent} approaches and representations, such as Neural Radiance Fields (NeRFs)~\cite{nerf,martinbrualla2020nerfw} and 3D Gaussian Splats (3DGS)~\cite{3dgs}, as well as from classic works in view-dependent geometry~\cite{vdg}. Namely, our core idea is to design a \textit{2D-driven} view-dependent deformation scheme, applicable to both 3D Gaussian Splats or triangle meshes.  

Our approach, illustrated in Figure~\ref{fig:interpolation},  enables a user to view the 3D object from a specific view direction $v_i$, generate a 2D mesh of the 2D image of the object from that view, and interact and deform it, thereby defining a deformation of the object $\phi_i$.  Subsequently, a number of these view-specific 2D deformations together define a \textit{deformation field} $f_v$ over the space of views $v$, i.e., a different 3D deformation is applied from any view direction, with a smooth, natural interpolation between the different views. 

We devise our novel approach for producing view-dependent 2D-based deformations of the object through our core observation, which is that we can define the deformation field that interpolates between the 2D deformations, without any coupling to the underlying geometry, nor between the different deformations (e.g., they do not need to share the same rig).  This enables us to design our method specifically for practical use, by targeting several critical properties: 1) Since the 2D deformations are not coupled with 3D geometry,  they can easily handle models with \textit{millions} of Gaussian Splats at interactive rates;  2) our scheme is not limited in the number and location of the selected keypoint views to interpolate, i.e., a user can perform multiple minute edit from any viewpoints they choose, to achieve an intricate result; 3) the deformations are \emph{composed} in a sequential, layer-like manner, i.e., each one operating on the deformed model produced by its predecessor, instead of simple blending between all of them as a linear weighted sum. This enables artists to build complex layers of deformations that modify previously deformed content, achieving detailed effect. Finally, we propose a method to apply the resulting deformation field to deform 3DGS/meshes - as far as we are aware, we are the first to propose a 2D-driven deformation technique for 3DGS. 

\begin{figure}
    \centering
    \includegraphics[width=\linewidth]{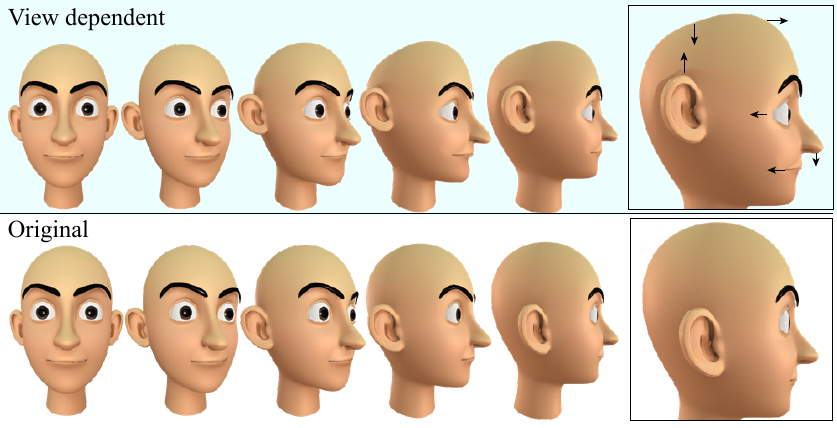}
    \caption{\textbf{view-dependent positioning of facial features.} Our method enables artists to design characters whose facial features change position and shape as the camera changes its viewpoint. In this example we emulate an edit similar to the one shown by the custom system designed for the Peanuts movie~\cite{peanuts}.}
    \label{fig:cartoon_head}
\end{figure}

Our method enables 2D artists to author 3D content that still exhibits the view-dependent qualities of the aforementioned 2D styles, while at the same time integrating into a general 3D framework, i.e., can be viewed as part of a scene, from all view directions. 
Figure~\ref{fig:teaser} exhibits such an example: at the top row, a 3D model of Rodin's sculpture ``The Thinker'' does not perfectly align with a 2D artist's sketch of the statue (left, in red). Our method enables a novice user to warp the 3D model from the chosen specific viewpoint on the left column, until it fits the sketch perfectly. Our approach then still enables treating this deformed model as a 3D object, interpolating between different deformations as the camera pans across it in a seamless manner, blending between the deformed view on the left and the original, undeformed view on the right.

We show the efficacy of our system through various experiments, such as editing of 3D models of realistic objects, recreating 3D versions of well-known art that violates consistent perspective, as well as cartoon-like results. We additionally show applications of this approach in various scenarios, such as fitting to sketches and caricatures,  selectively modifying proportions, and applying forced perspective. 

To summarize, our contributions are:
\begin{enumerate}
    \item We propose a method that ties between 2D edits from specific viewpoints to a continuous 3D deformation in a way that enables artists to produce complex view-dependent effects. 
    \item We propose a novel approach for interpolating between 2D deformations after lifting them to 3D, which enables both compositionality (each deformation is applied to the result of applying all previous deformations, as ``layers'', as opposed to a naive linear blending of all of them), as well as agnosticism to the representation of the underlying 2D deformations.
    \item We devise a novel, straightforward approach to apply general 3D deformations to 3D Gaussian Splats.
    \item We propose a simple technique to author the necessary 2D deformations of a given object from a specific viewpoint.
\end{enumerate}

%% file: sections/2_related_works.tex
\begin{figure}
    \centering
    \includegraphics[width=\linewidth]{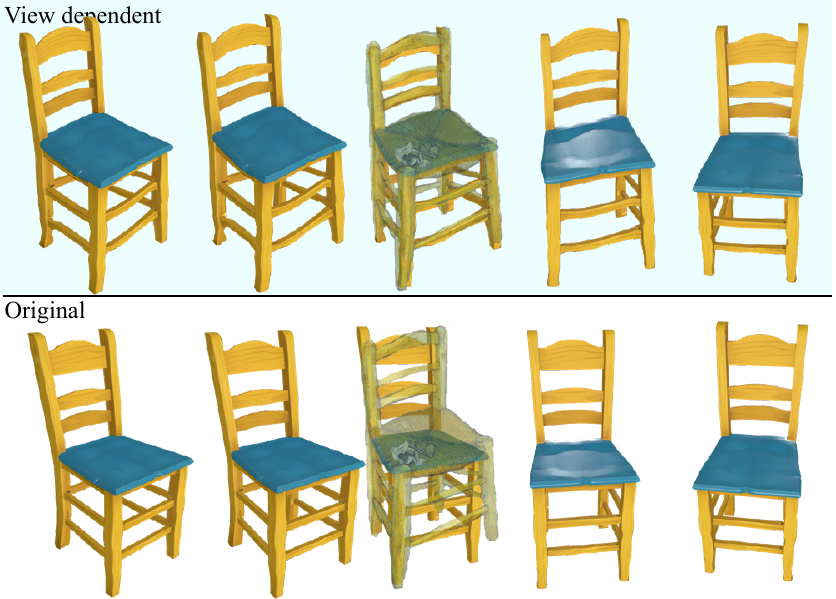}
    \caption{\textbf{Reproducing Van Gogh's chair as a view-dependent 3D model.} Our method enables a user to fit a 3D chair to Van Gogh's painting,  breaking perspective rules while still appearing correct from other views. }
    \label{fig:van_gogh_chair}
\end{figure}
\section{Revisiting View-Dependent Geometry}
\label{ss:vdg}
Our work is deeply inspired by View-Dependent Geometry (VDG)~\cite{vdg}, which was, as far as we know, the first work to propose to modify geometry in a view-dependent manner. Unfortunately, in the $25$ years that have passed since that work, its approach hasn't been picked up for practical application, in part due to several limitations, discussed below. In a sense, our work revisits the core concept of that work, but uses a rather different approach that in turn enables us to provide a much more expressive, intuitive, flexible and practical method to author view-dependent 3D content, in hope of reviving interest in view-dependent 3D modeling.

Namely, VDG consider view-dependent \emph{geometry}, i.e., each keypoint view holds a different positioning of the mesh's vertices $V_i$, which are then blended based on viewpoint. In contrast, we consider view-dependent \emph{deformations} of the 3D model, i.e., blending between \textit{functions} that deform 3D space, $\Phi_i:\mathbb{R}^3\to\mathbb{R}^3$. By that we can decouple the deformation from the underlying 3D model, and treat only the deformations as the view-dependent quantity. Furthermore, we propose a non-linear blending between the viewpoints, by composing the deformations, which also supports blending between any number of views, and controlling the range of effect of each deformation. This enables an artist to compose multiple small edits from different views, creating an intricate and subtle effect. In contrast, VDG uses linear blending between exactly three keypoint views at a time, severely limiting the expressivity of the method. Lastly, VDG design a method targeting deformations of 3D meshes, while we propose an additional extension of our method to 3D Gaussian Splats. All the above leads to three critical properties that distinguish our work from VDG:

\begin{figure}
    \centering
    \includegraphics[width=\linewidth]{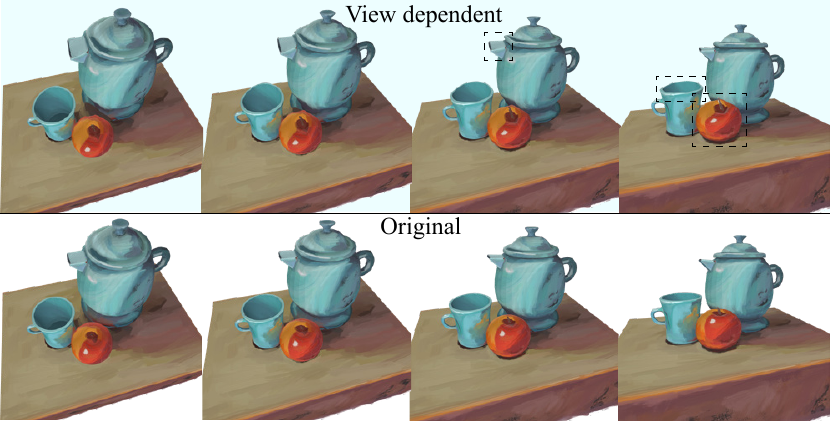}
    \caption{\textbf{Still life in the style of \cezanne{}.} We emulate \cezanne{}'s still life paintings,  famous as  one of the first examples of objects presented in multiple perspectives at the same time.}
    \label{fig:still_life}
\end{figure}
\begin{enumerate}
    \item \textbf{Significantly-greater expressivity.} The agnosticism to the underlying geometry, as well as to the type of deformation in each view, enables us to use different representations for the deformation from each viewpoint. Namely, we propose to use 2D ``deformation rigs'' for each viewpoint (see Figure~\ref{fig:2D_deform}), which provide much greater accuracy and flexibility than one single 3D rig.  Furthermore, considering deformations as functions enables us to define a \emph{compositional} blending of deformations in a sequential manner, where the user can apply another deformation on top of the already-deformed model, leading to complex effects (see Figure~\ref{fig:comparison_vdg_ours}), and also blend between \emph{any number of deformations} to achieve complex interactions (see Figure~\ref{fig:many_deformations}). 
    \item \textbf{Practical efficiency and applicability to modern 3D representations.} Instead of making another copy of a high resolution 3D model for each viewpoint, we only need to make a copy of the \textit{deformation}, which has significantly less degrees of freedom, thereby enabling practical application on high resolution models and Gaussian Splats. As shown in Figure~\ref{fig:graph}, VDG cannot be efficiently run on modern 3DGS models. We also propose a method to apply our deformation to 3DGS, which VDG does not consider.
    \item \textbf{Intuitive 2D representation.} Our method enables us to define each keypoint deformation as a 2D warp in the image plane, thereby providing a simple, intuitive 2D approach to interact and modify the appearance of the model (see Figure~\ref{fig:2D_deform}). 
\end{enumerate}

\section{Related Works}
\paragraph*{View-Dependent appearance.}
 The concept of modifying appearance from specific view points is a subfield of non-photorealistic rendering~\cite{gooch2001non} which has been well researched within computer graphics. 
 Many works followed up on view-dependent geometry~\cite{vdg}, such as works authoring spatial key frames which define a deformation across space~\cite{igarashi2006spatial}, again relying on 3D modeling and not supporting 3DGS. Other works focused on structures that change their 2D appearance as optical illusions and impossible structures~\cite{sela2007generation,li2024possibleimpossibles}, different types of non-linear projections~\cite{coleman2004ryan,yang2005nonlinear,sudarsanam2008non}, or notions of 3D ``canvases''~\cite{schmid2011overcoat} where 3D strokes do not directly map to 3D consistent objects. 
Another highly-relevant approach is 2.5D animation~\cite{rivers2010cartoon}, which uses ``billboards'' - 2D textured meshes floating in 3D space and directed towards the camera - in order to achieve a cartoonish 3D effect, which was extended to include view-dependent deformations later on~\cite{fukusato2022view}, to account for cartoonish effects that cannot be achieved via standard 2.5D animation. However, this method is designed specifically for the billboard representation of~\cite{rivers2010cartoon} and cannot generalize neither to 3DGS nor to 3D meshes. 
Lastly, we note that the term ``view-dependence'' is often mentioned in other contexts than modifying a 3D model, e.g., accounting for occlusions~\cite{tong2015view}, which is in essence a very different research area than the one discussed herein.

% --------------------------------- House  -----------------------------------------%
\begin{figure}
    \centering
    \includegraphics[width=\linewidth]{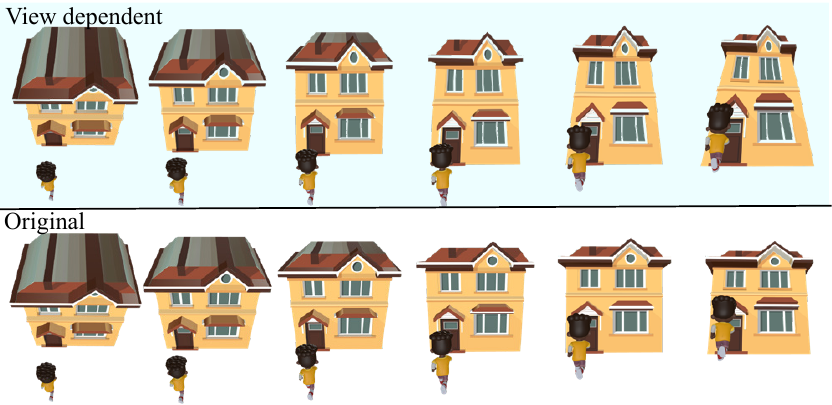}
    \caption{\textbf{Reproducing forced perspective effects.} A house can easily be made to tower over a kid, by making it more trapezoidal as the camera descends on the scene.}
    \label{fig:house}
\end{figure}

\paragraph*{Deformations and 3D modeling in computer graphics.}

Deformations play a crucial role in computer graphics, namely in applications such as modeling~\cite{sorkine2004laplacian,yu2004mesh}, and animation of, e.g., human faces~\cite{seol2011}, bodies~\cite{jacobson2012fast}, or elastic objects~\cite{de2017regularized,modi2024elastic}.  Additionally, deformations also stand behind important methods in 3D vision, such as registration~\cite{amberg2007optimal} and tracking~\cite{wang2023tracking}. Often,  the research of a specific deformation techniques is coupled with one of these target applications, or several of them. Deformations are often controlled by a set of controllers such as cages~\cite{ju2005mean}, bones~\cite{wang2015linear}, and points, which define the deformation of a 3D model via, e.g., linear blend skinning~\cite{skinningcourse:2014}, quaternions~\cite{kavan2008geometric}, or specialized coordinates~\cite{lipman2008green}. Other methods propose to deform the model by techniques such as computing the least-distorting deformation~\cite{sorkine2007rigid,botsch2006primo}, computing deformations that hold special mathematical properties such as prescribed curvatures~\cite{crane2011spin}, or through machine  learning~\cite{jakab2021keypoint,aigerman2022neural}. Our method uses a well-known mesh-deformation method~\cite{bbw} in order to deform 3D Gaussians.

\begin{figure}
    \centering
    \includegraphics[width=\linewidth]{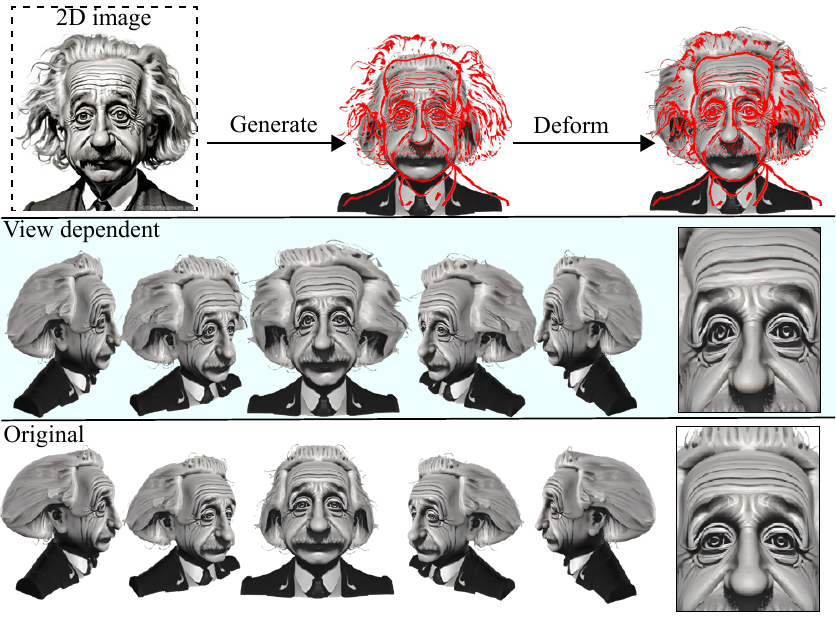}
    \caption{\textbf{Use within generative processes.} A 2D caricature of Einstein (top left) is used as input to a generative technique to generate a 3D model, which does not exactly match the input image (top middle). Our method enables a novice user to manually repair the model by warping the eyes, brows, nose, forehead and hair to fit exactly the original input (top right), but without any visible artifacts when observing it from other views.   }
    \label{fig:caricature}
\end{figure}

\paragraph*{Interfaces for 2D-based 3D modeling.}
 Many previous works have focused on ways to provide 2D interfaces for 3D content creation, in order to bridge between 2D image creation and 3D modeling. One prominent example is sketch-based interfaces for \emph{generating} 3D models~\cite{igarashi2006teddy,nealen2007fibermesh,gryaditskaya2020lifting,hahnlein2022symmetry,zhang2022creatureshop,brodt2022sketch2pose,puhachov2023reconstruction}. Similarly, ML-based methods for 3D generation can either be guided by sketches~\cite{delanoy20183d,li2018robust,zhong2020towards,zhang2021sketch2model,guillard2021sketch2mesh,binninger2024sens}, or images~\cite{kanazawa2016learning,wu2020unsupervised}. Closer to our work, other methods propose 2D sketch-based guidance for \textit{deformation}~\cite{nealen2005sketch,kho2005sketching,zimmermann2008sketching,Kraevoy:09}, however these focus on interpreting 2D strokes as gestures for deformation, do not yield view-dependent deformations, and additionally cannot be directly employed for deforming 3D Gaussian Splats.

\paragraph*{Deforming Gaussian Splats.}

3D Gaussian Splats (3DGS)~\cite{3dgs} have only recently emerged as a highly promising representation of 3D objects and scenes. Modification of them has thus far mainly focused on automatic diffusion-based methods using text techniques~\cite{wang2024gaussianeditor,chen2024gaussianeditor,chen2024dge,wu2024gaussctrl}. Specifically for deformations, they have arisen more in animation contexts, again automatically driven by a video diffuser~\cite{ling2024align} or a reference video~\cite{ren2023dreamgaussian4d}.
More specific to this work, considering deformations of 3DGS, PhysGaussians~\cite{xie2024physgaussian} propose a method to make 3DGS behave as deformable elastic objects, however the focus of the authors is on plausible physical simulation. Others~\cite{guedon2024sugar, gao2024mesh,jiang2024vr} bind 3DGS to a mesh that can be deformed, targeting rigid and articulated motions but less so free-form editing, while other approaches propose sparse controls~\cite{huang2024sc}. {Recently, sketch-guidance was used to deform 3DGS by manipulating a 3D cage~\cite{xie2024sketchguidedcagebased3dgaussian}.} In sum, these techniques are not directly applicable to the task we aim to achieve: 2D-based, view-dependent deformations of 3D objects. 

%% file: sections/3_method.tex
\section{Method}
We next lay out the different components of our method, starting with how we define a 3D deformation field from given 2D deformations from different viewpoints; how do we apply this deformation field on Gaussian Splats and meshes; and, how do we enable users to author the 2D deformations.

\subsection{View-dependent deformation fields from 2D deformation keyframes}
\begin{figure}
    \centering
    \includegraphics[width=\linewidth]{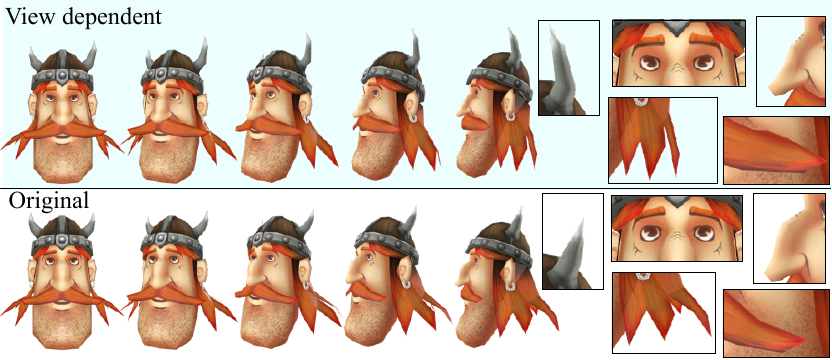}
    \caption{\textbf{Touching-up small details.} Our method is applicable for view-dependent editing of details at any scale: we modify the appearance of a viking model, represented as a 3D triangle mesh. While the model appears close to identical from the front, we modify its profile to give it a more refined nose, bigger mustache, less spiky hair, and an angled horn to his helmet.}
    \label{fig:viking}
\end{figure}
During the deformation process, the user selects a viewpoint $v_i$, i.e., a point on the sphere, $v_i\in S(2)$. We choose to represent the viewpoints in spherical coordinates (azimuth $v^\theta$ and polar $v^\eta$), as we find these represent well the human perceptual approach to view direction. The user then defines a 2D deformation of the image of the object from that viewpoint, $\phi_i:\mathbb{R}^2\to\mathbb{R}^2$ (see Figure~\ref{fig:interpolation} and Figure~\ref{fig:2D_deform}). 

Our goal is to define a deformation field, i.e., a function $f_v(p)$, receiving as input \textit{any} 3D point $p\in\mathbb{R}^3$ (not necessarily a centroid of a GS3D nor a vertex of a mesh), and \textit{any} view direction $v\in S(2)$ (not necessarily one of the chosen viewpoints). The output of $f_v(p)$ is the deformed 3D position of the input point,  that is, $f_v:\mathbb{R}^3\times S(2)\to\mathbb{R}^3$. 

This $f_v$ must satisfy two properties: 

\begin{enumerate}
    \item It must exactly align with each 2D deformation selected by the user, when viewed from the corresponding view direction, i.e., satisfy the following relation \begin{equation}
    \label{eq:congruent}
        \pi_i(f_{v_i}(p)) = \phi_i(\pi_i(p)),
    \end{equation} with $\pi_i$ being a projection into the 2D plane viewed from $v_i$. 
    \item $f_v(p)$ must be a smooth function in $v$, i.e., gradually change the resulting deformation as the viewpoint is changed. 
\end{enumerate}
In order to achieve this, we first design a way to lift 2D deformations to 3D ones, then devise a way to interpolate between these deformations in a compositional manner.
\paragraph*{Lifting 2D deformations to 3D.}
To begin, for each given 2D deformation $\phi$ and view direction $v$, we define a 3D deformation $\Phi$ that exactly agrees with $\phi$  when  viewed from the view direction $v$, i.e., $\Phi$ satisfies Equation~\eqref{eq:congruent}.
Towards that goal, consider how a 3D point $p=\left[x,y,z\right]\in\mathbb{R}^3$ is mapped to screen coordinates: first, $p$'s 3D position in the local camera's coordinates from a specific viewpoint $v$ is defined by
\begin{equation}
    \psi(p) = R_i  p + t_i,
\end{equation}
where $R_i\in SO(3)$ is a rotation matrix and $t_i\in\mathbb{R}^3$ is a translation vector. For ease of notation, we will assume without loss of generality that we have already applied this transformation, and $p$ is already represented in local camera coordinates.

% --------------------------------- Castle  -----------------------------------------%
\begin{figure}
    \centering
    \includegraphics[width=\linewidth]{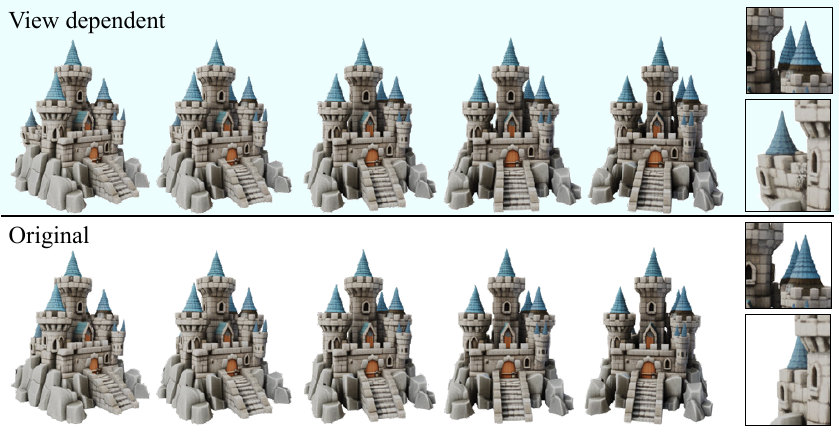}
    \caption{\textbf{Avoiding occlusions.} Specific viewpoints may occlude some desired parts of a model (e.g., the rooks of the castle) - our method enables slightly shifting them so that they appear visible from desired views.}
    \label{fig:castle}
\end{figure}

The camera itself has (global, view \textbf{in}dependent) parameters defining its field of view, given by a matrix 
\begin{equation}
K =
\left[
\begin{array}{ccc}
a_x & 0 & c_x \\
0 & a_y & c_y \\
0 & 0 & 1
\end{array}
\right],
\end{equation}
which is used to define the perspective projection onto the screen:
\begin{equation}
    \pi( p) = K \cdot p \cdot {z}^{-1}.
\end{equation}

Hence, we need to devise a 3D deformation $\Phi$ s.t. when it is projected onto the screen, it agrees with the 2D deformation $\phi$, i.e.,  satisfies Equation~\eqref{eq:congruent}. We make the natural choice of moving the 3D point in the plane parallel to the 2D projection plane, to the exact location such that it will be projected to the desired 2D point, while maintaining its distance to this plane, i.e., 
the desired deformation in camera coordinates is 
\begin{equation}
    \Phi(p) =K^{-1}\cdot\phi(\pi(p))\cdot z.
\end{equation}
Note that in order to obtain the final 3D deformation $\Phi$, we only require the ability to \emph{evaluate} $\phi$, but can treat it as a black box otherwise - this will enable us to interpolate between different deformations that do not share a view plane nor the same representation (namely, each defined by a different 2D mesh).
%%%
Finally, to return to global world coordinates we simply perform the inverse transformation on the deformed point: $\psi^{-1}\left(\Phi\left(p\right)\right)$. See Algorithm~\ref{alg:deformation3d} in Appendix~\ref{ap:algo}

\paragraph*{View-dependent interpolation of the deformations.}

Once we have  3D deformations $\Phi_1,...,\Phi_n$ corresponding to viewpoints $v_1,...,v_i$, we define a view-dependent \textit{interpolation} between them (see Figure~\ref{fig:interpolation}). 

Each viewpoint $v_i$ is assigned a \textit{basis function} $B_i$, used to weigh the deformation $\Phi_i$ w.r.t. to all the other deformations, so that it has a localized effect. Namely, $B_i$ receives a view point $v$ as input and outputs a scalar, which is $1$ when $v = v_i$ and monotonically decays to zero as the viewpoint $v$ becomes farther from $v_i$, using the distance of their angles in polar coordinates:

\begin{equation}
    B_i(v) = e^{ -\sigma^i_1(v^\theta-v_i^\theta)^2 -  \sigma^i_2( v^\eta-v_i^\eta)^2},
\end{equation}
where $\sigma^i_1,\sigma^i_2$ are user-chosen parameters for that specific view, that control how fast the deformation drops to have no effect as the viewpoint changes, and the subtractions are performed in a periodic manner w.r.t. angles.

\begin{figure}
    \centering
    \includegraphics[width=\linewidth]{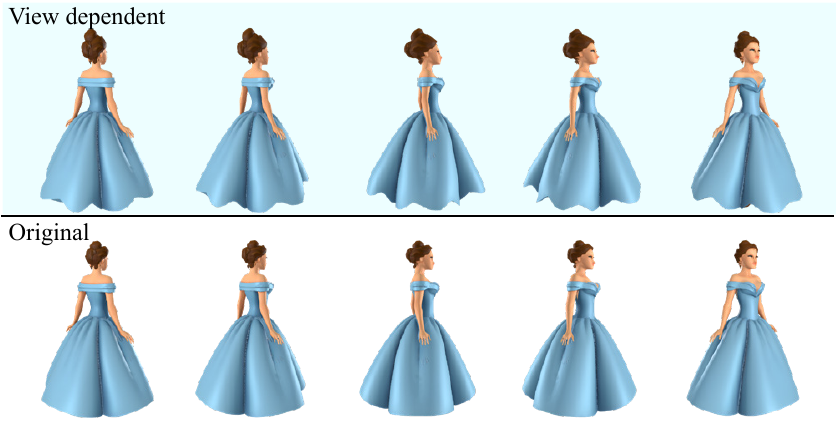}
    \caption{\textbf{Editing cloth.} Pieces of fabric often appear either non-realistic or otherwise do not align with artistic desires, such as the dress lacking in folds. Our method enables editing the cloth in second, without need to worry about physical plausibility, as the 2D edits morph into each other naturally as the dress is rotated.}
    \label{fig:blue_dress}
\end{figure}
\begin{figure*}
    \centering
    \includegraphics[width=\linewidth]{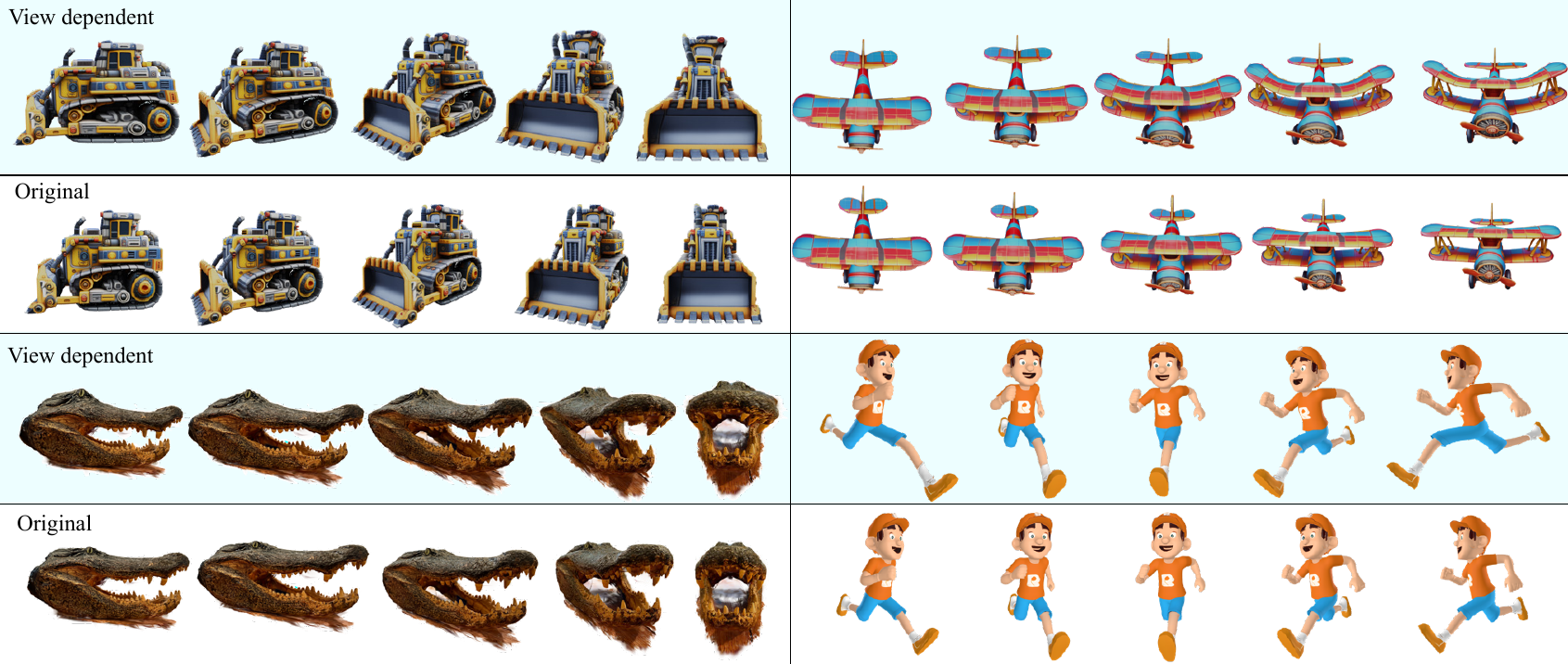}
    \caption{\textbf{Breathing subtle dynamism into 3D models by simple, novice-level edits.} Subtle edits of some proportions and positions from specific views can be almost undetected by an unaware viewer, disappearing with viewpoint changes, however still  have a great effect.}
    \label{fig:exaggerated}
\end{figure*}
Previous interpolation schemes, such as View-Dependent Geometry~\cite{vdg}, use a piecewise-linear basis to \emph{sum} a subset of deformations, weighted by basis functions, to blend them. However, this approach suffers from the common pitfalls of linear blending (see comparison in Figure~\ref{fig:comparison_vdg_ours}). Instead, we aim for a "layered" compositionality, {where deformations are applied sequentially, each modifying the already-deformed model rather than being averaged together. This ensures that successive deformations accumulate progressively, preserving their distinct effects without overriding previous transformations.} Towards this end, we define a simple recursive procedure, defining for each of the $n$ static deformations $\Phi_k$ a corresponding view-dependent deformation $D_k$, by blending between $\Phi_i$ and all previous deformations:

\begin{equation}
\label{eq:interpolation_point}
\begin{split}
    D_k(p, v) = B_k(v) \cdot \Phi_k(D_{k-1}&(p)) + (1 - B_k(v)) \cdot D_{k-1}(p), \\
    D_1(p, v) &= \Phi_1(p).
\end{split}
\end{equation}

Finally, the view-dependent deformation field is given by  

\begin{equation}
    f_v(p) = D_n(p, v).
\end{equation}
This entire computation is summed up in Algorithm~\ref{alg:interpolation}.

\subsection{Applying the deformation field to 3D Gaussian splats}
\paragraph*{3D Gaussian Splats.} 3DGS are defined by a set of $k$ 3D Gaussians, each with a mean $\mu_i$ and variance $\Sigma_i$, s.t. the $i$'th Gaussian is defined as 

\begin{equation}
\label{eq:gaussian}
    G_i(p)=e^{(p-\mu_i)^T\Sigma_i^{-1}(p-\mu_i)},
\end{equation}
which together with color parameters represent a 3D scene. They are rendered onto the screen by projecting them onto the 2D view plane (``splatting''), leading to a highly-efficient rendering process, as well as fast fitting to given images and geometry. 

\paragraph*{Applying the deformation field.} Given a deformation field $f_v(p)$, we wish to deform the Gaussian splats, observed from view direction $v$ (not necessarily one of the selected keypoint views $\{v_i\}$).

In order to account for the deformation, the correct mathematical formulation for a deformation of a Gaussian would be to "pull" the original Gaussian's values into deformed space, $\tilde G(p) = G(f_v(p))$, however this results in a function $\tilde G$ that is no longer a Gaussian, which in turn would prevent their use in 3DGS pipelines. 
Instead, in order to produce Gaussians, we propose to use the best affine approximation of the deformation, using a first-order Taylor expansion of $f_v(p)$ around the Gaussian's centroid. The Taylor expansion is $L(p) = f_v(\mu) + J (p-\mu) $, where $J$ is the \emph{jacobian} of $f_v$ at point $\mu$ (the jacobian could be directly computed using automatic differentiation, however we derive it directly by applying the chain rule to Equation~\eqref{eq:interpolation_point}, along with derivating $\Phi$). This affine approximation is a good approximation of $f_v$ for small-enough Gaussians, and most importantly,  \emph{does} deform Gaussians into Gaussians: plugging $L(p)$ instead of $p$ into Equation~\eqref{eq:gaussian}, it is immediate to deduce that the new mean and variance are

\begin{equation}
    \tilde \mu = f(\mu,v),
\end{equation}
and
\begin{equation}
    \tilde \Sigma = J^T\Sigma J. 
\end{equation}
When deforming the Gaussians we thus iterate over each $G_i$, and define its new deformed version $\tilde G_i$ using the formulae above. As far as we are aware, although straightforward, this approach has not been applied yet to 3DGS.
\begin{figure}
    \centering
    \includegraphics[width=\linewidth]{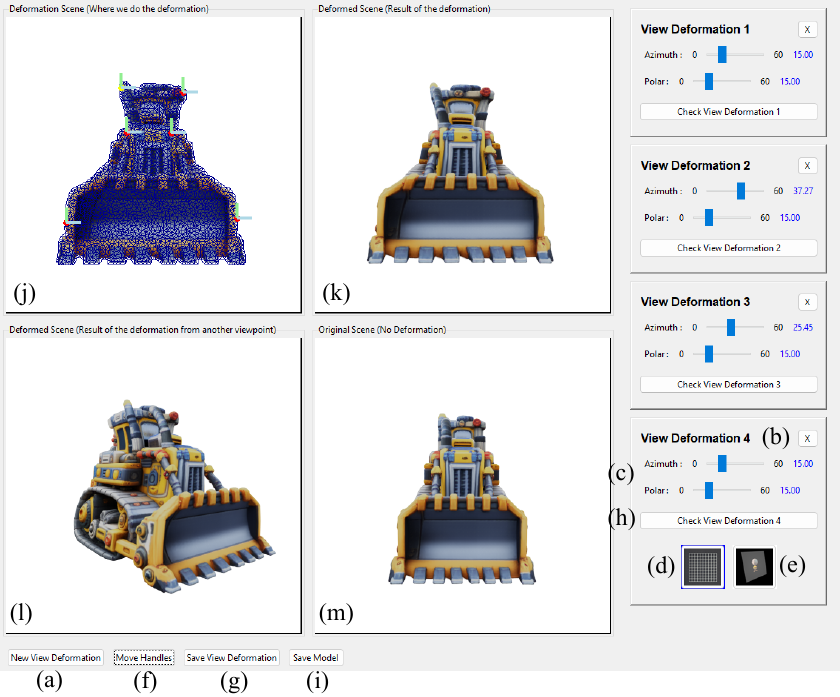}
    \caption{The user interface we use to author view-dependent deformations. See Appendix~\ref{ap:gui} for a full explanation.}
    \label{fig:gui}
\end{figure}
\paragraph*{Deforming 3D meshes.} Of course, the above method can also directly support deformation of 3D meshes (see Figure~\ref{fig:viking}), in which case we simply treat each vertex $V_i$ of the mesh as a Gaussian with mean $\mu_i$, while ignoring the notion of variance $\Sigma_i$.

\subsection{Authoring 2D deformations}
\label{ss:author}
 We next detail our approach to authoring the required 2D deformations, which interacts especially  well with the idea of view-dependent content.
However, note that our system is completely modular, in that the interpolation scheme is applicable to any type of 3D deformations $\Phi_i$, and additionally the 2D-to-3D lifting scheme can be applied to any type of 2D deformations $\phi_i$, enabling other types of 2D interactions and authoring. 

\paragraph*{2D deformation pipeline.} Our method progresses in several stages, as visualized in Figure~\ref{fig:2D_deform}:
\begin{enumerate}
    \item The user selects a viewpoint $v$ of the object (Figure~\ref{fig:2D_deform}, left). \item the object is rendered from that viewpoint, into a 2D raster image; The 2D raster image is then triangulated  using the algorithm implemented in Triangle~\cite{shewchuk1996triangle} into a 2D mesh (Figure~\ref{fig:2D_deform}, middle). 
    \item The user selects ``handles'' (vertices of the mesh, visualized as magenta circles in Figure~\ref{fig:2D_deform}) that they will then drag to deform the mesh; using Bounded Biharmonic Weights (BBW)~\cite{bbw}, we ``rig'' the mesh w.r.t. the handles so that moving the handles affects non-handle vertices of the mesh.
    \item The user interacts with the handles (Figure~\ref{fig:2D_deform}, middle, bottom) in order to deform the mesh until they are satisfied with the resulting deformation, at which point the result is stored as the 2D deformation $\phi_n$.
    \item Given the resulting deformation of the 2D mesh, we can move any given 2D point $q\in\mathbb{R}^2$, by finding the triangle it falls inside of in the undeformed mesh, computing barycentric coordinates with respect to it, and then placing the point in the barycenter of the deformed mesh. \martin{(see Algorithm~\ref{alg:deformation2d} in Appendix~\ref{ap:algo})}

\end{enumerate}

Lastly, we designed a simple GUI to author these deformations, shown in Figure~\ref{fig:gui}, and discussed in Appendix~\ref{ap:gui}.

\begin{figure}
    \centering
    \includegraphics[width=\linewidth]{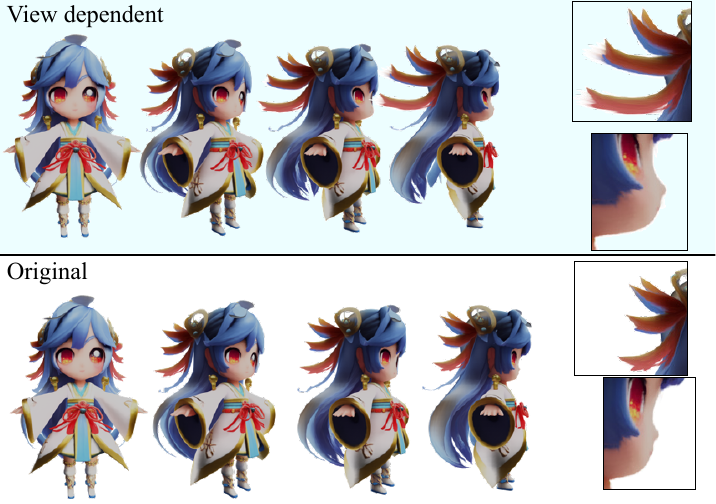}
    \caption{\textbf{Application to other aesthetic styles.} Our method can be applied in similar fashion to other styles than western cartoons and paintings.}
    \vspace{-10pt}
    \label{fig:anime_girl}
\end{figure}

\subsection{Implementation details}
We use GSplat~\cite{ye2024gsplatopensourcelibrarygaussian}  to render Gaussian Splats and PyRenderer for  meshes. Libigl~\cite{libigl} was used for the BBW~\cite{bbw} solver, and Pytorch~\cite{imambi2021pytorch} for computing the deformation. 
To create the 2D triangle mesh, we render the images at $400\times400$ resolution, and use OpenCV's contour detection to extract a 2D boundary, which is input to Triangle~\cite{shewchuk1996triangle} to obtain the triangulation. We call Triangle with a  minimal angle threshold of $32.5^\circ$, and a maximum area equivalent to $20$ pixels. 

\paragraph*{Timing.}    

When the user interacts with the GUI, the deformation computation (including all computations and rendering) runs in 25 FPS. The setup time before deformation is as follows: computing BBW weights: 100 miliseconds,
    triangulation: 33 miliseconds. All timings were conducted on a NVIDIA RTX4090 GPU, for a 3D model consisting of $750$K Gaussians.

\paragraph*{Models.} The bulldozer, plane, anime character, castle, and mushroom house are AI-generated 3DGS models , while the cartoon head, Einstein, chair, running boy, dress, and bunny head are generated mesh models. The alligator, tree, still life, and Rodin statue are 3DGS models, and the viking and house are meshes obtained from Sketchfab.

%% file: sections/4_experiments.tex
\section{Experiments}

We next detail various experiments we conducted to show the efficacy of our method in various scenarios of modeling 3D scenes, using various categories of 3D objects, obtained by different methods (artist-created, generative, real-life). 

\begin{figure}
    \centering
    \includegraphics[width=\linewidth]{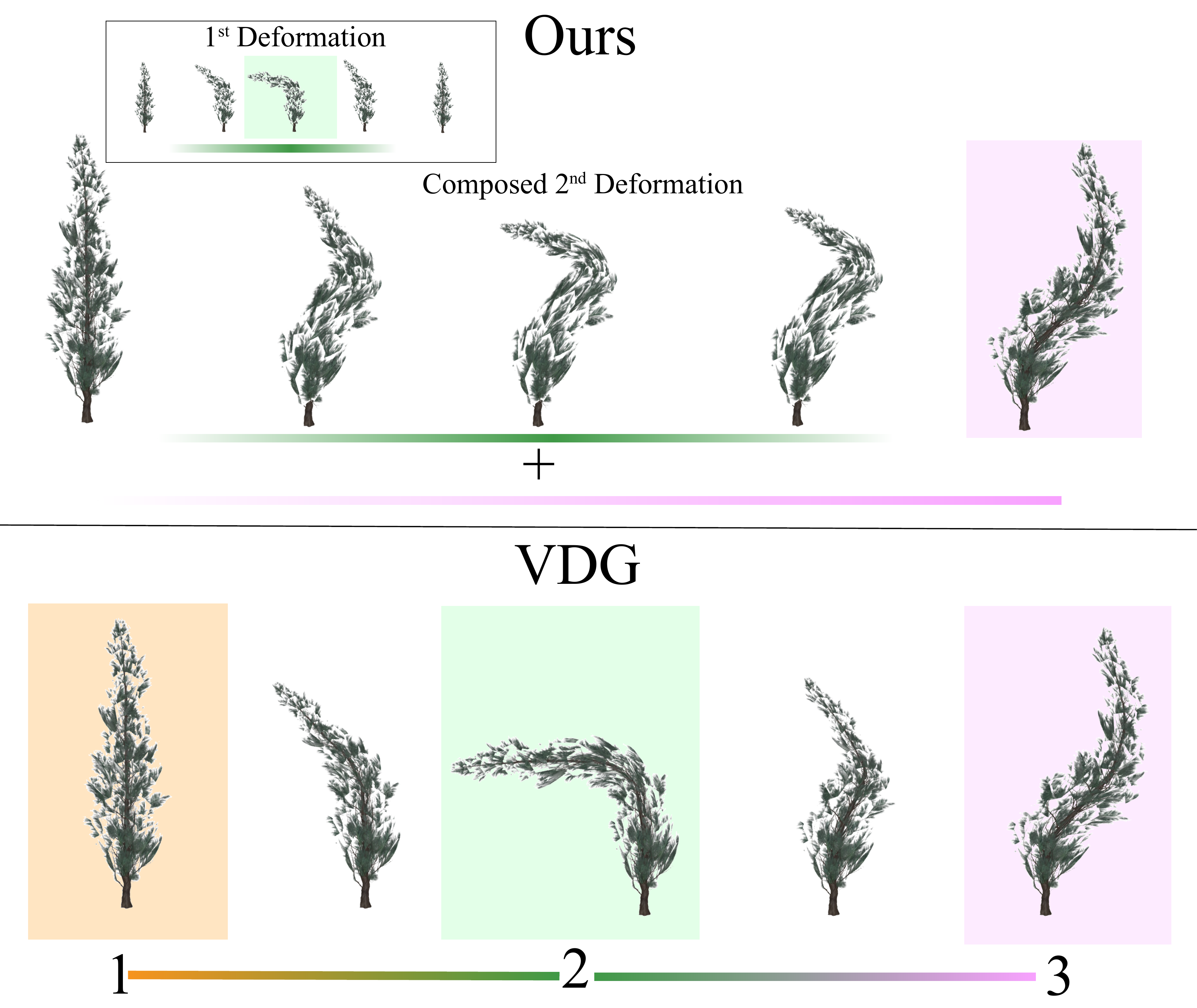}
    \caption{\textbf{Interpolation scheme.} Our method can compose the different deformations when interpolating, thus creating a smooth, complicated deformation of the tree. VDG~\cite{vdg} can only perform piecewise-linear deformation, hence linearly blends to the middle deformation, and then sharply transitions to interpolating to the rightmost deformation.}
    \label{fig:comparison_vdg_ours}
\end{figure}

\paragraph*{Artistic exaggeration.} Our method enables even pedestrian users, without much experience in graphics, to make subtle changes to 3D models to make them pop out and appear more dynamic. Figure \ref{fig:exaggerated} shows several such subtle edits that significantly impact the impression made by an object. The scaling of the bulldozer can only be achieved with view inconsistency, and leads to a more menacing result. Similarly, the upper jaw of the alligator is enlarged from the front to make it look more menacing. The plane appears more dynamic and ``midflight'', with a simple 2D edit that does not harm the appearance from an overhead view.  Likewise, the legs and arms of the runner are stretched to emphasize motion.

\paragraph*{Forced perspective effects.} Forced perspective is often used to give the impression that an object is larger than it really is, for example, by scaling proportions to make it appear as though the change in scale is due to differences in distance to the viewer and not by sheer differences in object size. Obviously, this effect is only feasible from specific views, lest the user sees that indeed the change in scale is not due to distance. Hence, this effect is perfectly suitable for our method, and we show such an example in Figure~\ref{fig:house}, where we make the house more trapezoidal as the camera descends, giving the appearance of the model being bigger than it actually is, towering over the human character.

\paragraph*{Occlusions.} It is often the case that the desired view of an object leads to unwanted occlusions. Figure~\ref{fig:castle} shows a castle that has a couple of its rooks hidden from specific viewpoints. Our method enables easily adjusting the position of the hidden rooks for those desired viewpoints, while still enabling a smooth transition to the other viewpoints without ruining the coherent geometric appearance of the object.

\paragraph*{Small-scale touch-ups.} Our method also provides artists the flexibility to apply localized touch-ups to the 3D model. For instance, Figure~\ref{fig:viking} illustrates a Viking head with specific adjustments. From the front view, we refine the eyes to enhance expression. From side views, we adjust the nose, the mustache, the helmet's horn, and the hair to achieve a thinner silhouette. Similarly, in Figure~\ref{fig:blue_dress}, we demonstrate how our method allows artists to refine a character's dress. In order to make the dress appear more lifelike, our 2D-based editing enables adding creases and folds to the cloth without requiring a 3D editing session, which would be demanding and possibly require cloth simulation. We additionally increase the volume of the hair to create a more stylized and expressive look. Our method can also be applied to anime characters, offering artists the ability to make stylistic adjustments with ease. In Figure~\ref{fig:anime_girl}, we showcase an anime-style  character edited from a side view, where we adjust the hair, stretching the ribbon, refining the character's silhouette, and adjusting the jawline. Figure~\ref{fig:cartoon_head} exhibits a model whose facial features shift when the viewpoint is changed, to represent a more-consistent style. This, in turn, is an emulation of the technique explained by the creators of the Peanuts movie~\cite{peanuts}. While they describe a custom rigging solution, our method provides a blackbox drop-in solution which can achieve similar effects, with a much simpler implementation.

\paragraph*{Fitting to 2D illustrations.} 3D models often do not exactly align with a target 2D illustration of them. Our method enables to fix this, by first rendering the object on top of the 2D illustration, and then warping the model in 2D until it exactly fits the target. Due to the view-dependent nature of our method, these edits do not need to be 3D consistent as they gradually resolve as the viewpoint is changed. Figure~\ref{fig:teaser} shows one such result, where we deformed a 3D model of Rodin's "The Thinker" to match an artist sketch (in red). The change is barely noticeable when panning around. Figure~\ref{fig:caricature} shows another use case of this approach, for adjusting AI-generated 3D models: we generate an image of a caricature of Albert Einstein using Stable Diffusion~\cite{rombach2022highresolutionimagesynthesislatent} (top left), and then generate a 3D model from this image (top middle, in gray, underneath the red contour of the input image). Evidently, the model does not align with the input image. A (novice) user alleviates this by using our method to manually deform the 3D model, refining it to stay true to the caricature it is supposed to match, by subtle changes to the eyes, brows and forehead, as well as the hair. These minute changes are easy to author in 2D, but would be much more labor-intensive in a 3D modeling software that would require 3D view consistency.

\begingroup
\setlength{\intextsep}{0pt}%
\setlength{\columnsep}{4pt}%
\paragraph*{Recreating famous skewed-perspective paintings.}

We were excited to apply our method to classic artistic \begin{wrapfigure}{r}{0.1\textwidth}
\centering
    \includegraphics[width=0.1\textwidth]{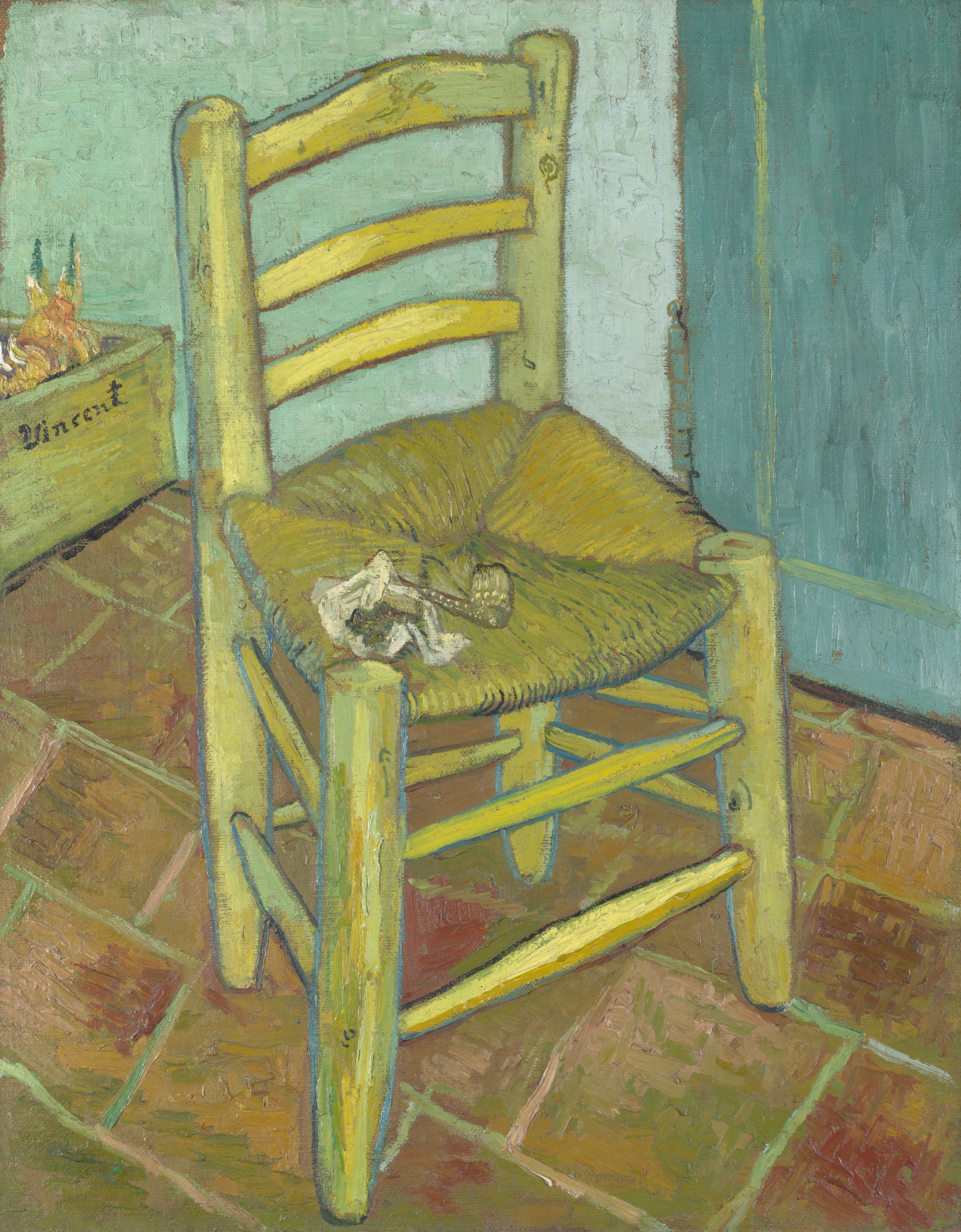}
\end{wrapfigure}paintings that are famous for incoherent perspective, and reproduce them as view-dependent objects. One such work is Van Gogh's Chair - we extracted a 3D model of it and then made it match the painting exactly. As Figure~\ref{fig:van_gogh_chair} shows,  when rotated, the chair still appears as a valid 3D object.

\begin{figure}
    \centering
    \includegraphics[width=\linewidth]{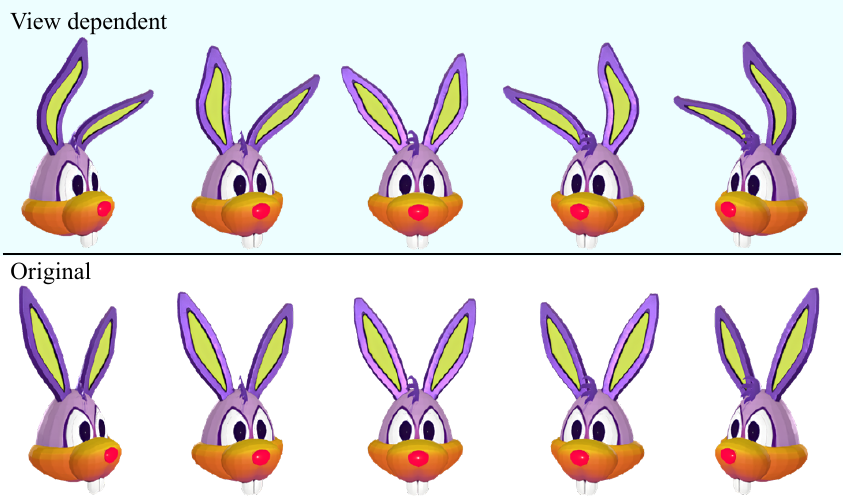}
    \caption{\textbf{Reproducing the main result from VDG~\cite{vdg}.} Our method can easily reproduce the main result from VDG (Figure 6 in their paper). }
    \label{fig:bunny}
\end{figure}

Similarly, in Figure~\ref{fig:still_life} we reproduce a scene in the style \begin{wrapfigure}{r}{0.13\textwidth}
    \includegraphics[width=0.13\textwidth]{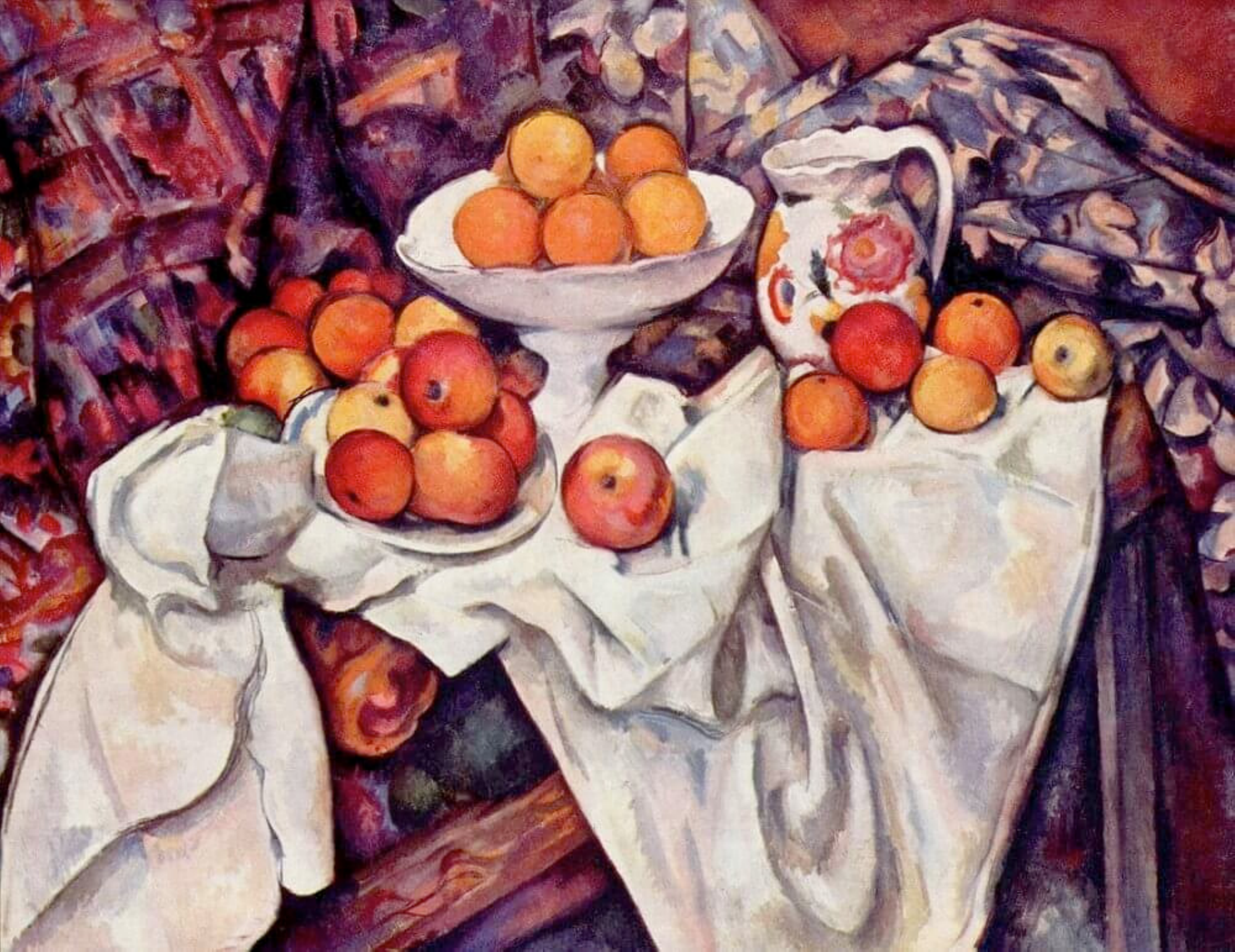}
\end{wrapfigure} of the still-life paintings of \cezanne{}, who is often attributed as the forefather of cubism, as those paintings often exhibited objects that were drawn from different perspectives. We achieve a similar effect by getting a still life scene, and then editing it to produce shifts in perspectives from different views (highlighted in dashed squares), such as making the fruit and mug face the camera when viewed from the front, and have the spout of the kettle follow the camera.

\endgroup

\begingroup
\setlength{\intextsep}{0pt}%
\setlength{\columnsep}{4pt}%

\subsection{Comparison to View-Dependent Geometry}   View-Dependent Geometry (VDG)~\cite{vdg} stands as the closest work to ours, and one of only a few that deal with modifying geometry conditioned on view direction. We now show experiments exhibiting the effects of the main  conceptual differences discussed in Section~\ref{ss:vdg}. 

\begin{wrapfigure}{r}{0.18\textwidth}
    \includegraphics[width=0.18\textwidth]{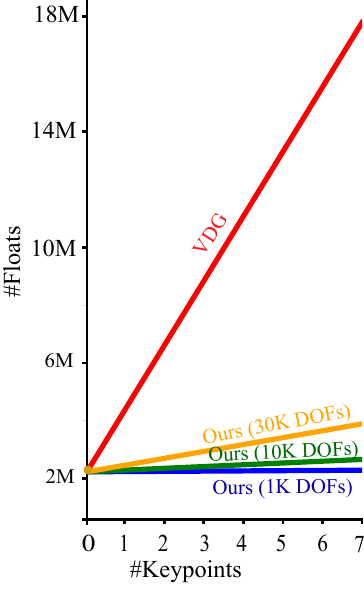}
    \vspace{-15pt}
    \caption{\textbf{Memory footprint.}}
    \label{fig:graph}
\end{wrapfigure}One main difference lies in the memory footprint. As explained before, VDG duplicates the 3D model for each additional keypoint view used. Hence, even if VDG were applicable to Gaussian Splats, the number of primitives would increase linearly with the number of keypoint views. In contrast, we only need to store the 2D deformation $\phi_i$ for each view, which amounts to storing the 2D mesh used for the deformation, which usually has around a $1000$ vertices. Figure~\ref{fig:graph} shows a graph of the number of floats consumed by VDG, per number of keypoint views, for a 3DGS model of $250K$ Splats. We also show our method's memory footprint, using three different 2D mesh resolutions (in practice, we never attain $30K$ vertices).

Another key difference lies in the interpolation schemes, see Figure~\ref{fig:comparison_vdg_ours} - VDG perform linear interpolation between the $3$ closest keypoint views, hence when interpolating around the tree, it in essence interpolates linearly between three configurations (highlighted with color frames) - the straight tree, which is interpolated linearly to the configuration in the middle, then another interpolation between the middle and the right. In contrast, our formulation enables \emph{composing} the deformations, starting with the same one used by VDG (top, left), but then composing it with another deformation (middle row), to create a smoother and more complicated deformation between leftmost and rightmost.  

Lastly, VDG can only interpolate between three keypoints at a time, and is limited in the number of total keypoints it can represent. In Figure~\ref{fig:bunny} we show that our method easily recreates the main result from Figure 6 in VDG (one of only a handful shown in that paper). To achieve a direct comparison, we use the image of the bunny from their paper and generate a similar 3D model. We then reproduce the exact view-dependent deformation from that paper. Furthermore, in Figure~\ref{fig:many_deformations}, we show an example in which we deform the bunny into a large amount of keypoint deformations (each keypoint view visualized as a point on the sphere), an example of a result that is strictly unattainable with VDG.
\endgroup

%% file: sections/5_conclusion.tex
\section{Conclusion}

The experiments above confirm the ability of our approach to author 3D assets which can account for view-dependent local 2D edits to achieve various expressive and artistic desires, exceeding what was achievable with previous techniques. We are excited to extend our GUI into a fully-fleshed application, which will enable many other types of edits and interactions, such as other types of rigs (bones, cages), or incorporating additional 3D deformation tools, such as a local 3D rotation, to, e.g., make a pair of bunny ears always follow the camera. Another important addition is the ability to add symmetric deformations, to, e.g., move both ears in the same manner - these are rather straightforward additions to our existing method, although we note that they were not necessary to achieve the expressive results shown in this paper.

\begin{figure}
    \centering
    \includegraphics[width=\linewidth]{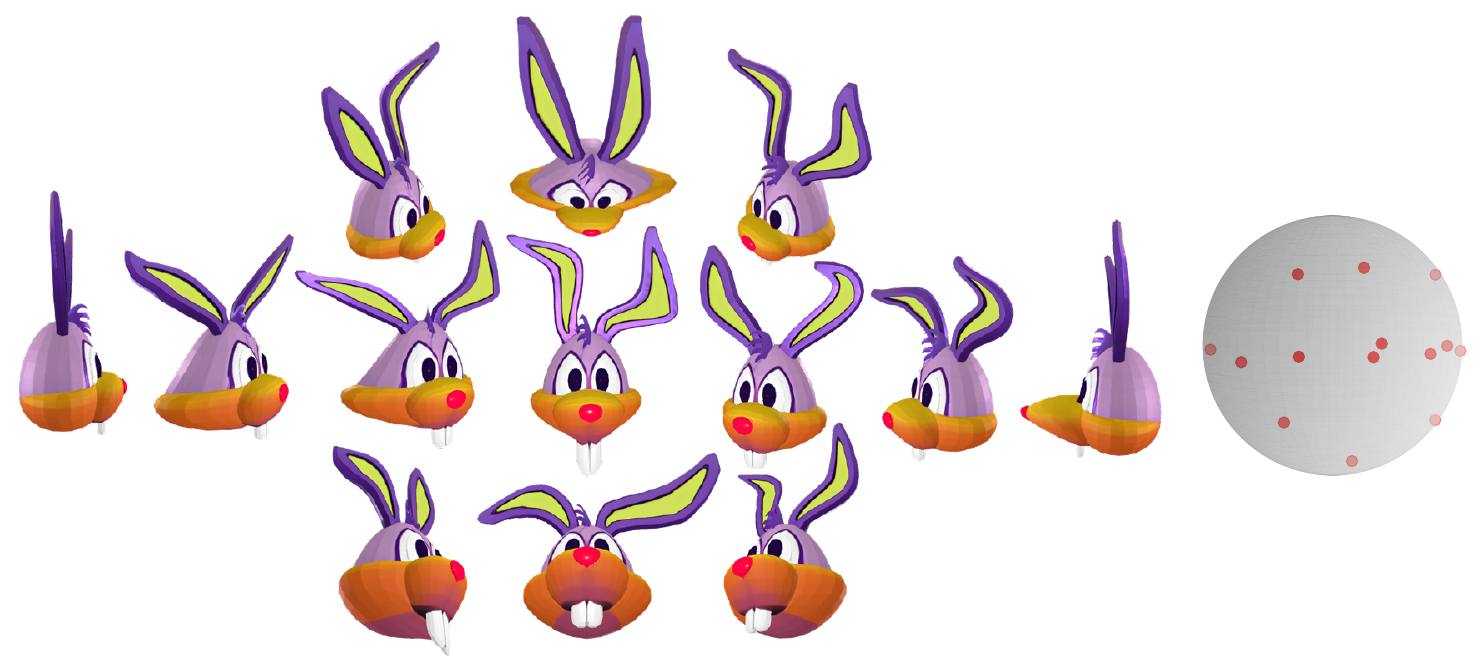}
    \caption{\textbf{Interpolating between a large number of keypoint views.} Our method supports any number of desired keypoint views, and can interpolate between all of them simultaneously, if desired. This is unattainable through VDG~\cite{vdg}.}
    \label{fig:many_deformations}
\end{figure}

While our method provides intuitive controls for view-dependent editing, it does hold limitations: first, we do not modify the number of Gaussians, and in case the 3D model contains large Gaussians, a deformation may cause them to be separated or misaligned, resulting in visible artifacts. This can be avoided by incorporating a splitting technique as in the original 3DGS paper~\cite{3dgs}, however, in practice this problem was not prevalent for the models we used. Second, we note that our technique may prove to be, at some times, too restrictive. For example, in some cases two parts of an object are close enough to be considered connected when rendered at low resolution, and as a result the 2D mesh used for the deformation will not provide the means to separate those two parts. 

We believe we have only scratched the surface of what is feasible using conditioned deformations. The most important frontier is of course animation. We focused on static objects as a first attempt at devising this method, however we believe our method can be easily adapted to interact with rigged characters, to enable view-dependent assets that can be controlled and animated - nonetheless, this requires further research beyond on the scope of this work, and we target it as a followup. Furthermore, we believe that we can extend our approach to not only be conditioned on viewpoints, but on the actual motion of the character, By that, we hope that we will be able to provide tools for a visual language built around movement,  similarly to how this work provides tools for a visual language built around viewpoints.

\paragraph*{Acknowledgments.} This paper was conceived at the \textit{2024 Bellairs Workshop on Computer Animation}, organized by Prof. Paul Kry, along with invaluable discussions with Prof. Maneesh Agrawala, Prof. Alec Jacobson, and Prof. David I.W. Levin. This work was funded through the Natural Sciences and Engineering Research Council of Canada (NSERC) Discovery grant ``Practical Neural Geometry Processing'', as well as an Adobe gift.

%% file: sections/6_gui.tex
\section{User interface for authoring deformations.} 
\label{ap:gui}

We devised a proof-of-concept, minimal GUI, which provides the critical realtime feedback necessary to enable the user to intuitively control and author deformations, as shown in Figure~\ref{fig:gui}. This GUI can be extended and improved, as our focus in this paper was not on the user experience but on the core idea of devising a modern computational approach to perform view-dependent deformations based on pure 2D deformations. 

The user can rotate the 3D model displayed in the GUI until finding a desired view point. Then they press on \myenum for logging the current view as $v_n$ and initiating a deformation process for generating $\phi_n$. This generates a new view deformation panel \myenum - one for each deformation $\phi_i$. This panel has an azimuth and polar sliders \myenum that control $\sigma_1,\sigma_2$ and affect how far from the chosen view will the deformation take effect. Clicking on \myenum then generates a 2D mesh on top of the 2D render of the object. We additionally provide the ability to choose a cutting plane \myenum which displays only the part of the model beyond some z value, in order to apply the deformation selectively. The user presses \myenum to compute the BBW~\cite{bbw} rig and initiate the interactive deformation process. When they are done, they can save the current deformation \myenum and progress to the next desired viewpoint. They can also orient the camera to the current viewpoint by pressing \myenum. When the process is complete, the model can be saved \myenum.

We display four renders of the model simultaneously: \myenum provides the interface to deform the model via the 2D mesh (see also Figure~\ref{fig:2D_deform}) - the user can click on the mesh's vertices to designate them as handles, as well as drag or rotate them; \myenum displays the same view, without the mesh and mouse cursor so that the user can view the result, unoccluded; \myenum displays the deformed model, from any other chosen view by the user; \myenum displays the undeformed model, for comparison. 

%% file: sections/7_pseudo_code.tex
\section{Deformation Algorithms}
\label{ap:algo}

%------------------------ Deformation process -----------------------------%
In the pseudo-code presented in Algorithm~\ref{alg:deformation_process}, $P$ represents the vertices for a 3D mesh and the means for a 3DGS, $\Sigma$ represents the covariance matrices for 3DGS, $\{\Phi_i\}$ is the set of 3D deformations, and $v$ is the current viewpoint.

\begin{algorithm}[H]
\caption{Deformation Process}
\label{alg:deformation_process}
\begin{algorithmic}[1]
    \State \textbf{Input:}  
    \State \quad \textbf{3D Model:} Mesh $(P)$ or 3DGS $(P, \Sigma)$ 
    \State \quad \textbf{Deformations:} \( \{\Phi_i\} \)
    \State \quad \textbf{Viewpoint:} v
    
    \State \textbf{Output:}  
    \State \quad \textbf{Deformed 3D Model:} Mesh $(P', F)$ or 3DGS $(P', \Sigma')$

    \State $n \gets length(\{\Phi_i\})$

     \For{each point $p_i \in P$}
        \State $p_i', J_i \gets  \textbf{D}(\{\Phi_i\}, p_i, v, n)$
        \If{3D model is a 3DGS}
            \State $\Sigma_i' \gets J_i \Sigma_i J_i$
        \EndIf
    \EndFor

    \State \Return $P'$ \textbf{or} $P', \Sigma'$ \text{if \textbf{3D model} is a 3DGS}
\end{algorithmic}
\end{algorithm}

%------------------------ Recursive Function -----------------------------%
In the pseudo-code presented in Algorithm~\ref{alg:interpolation}, $\{\Phi_i\}$ is the set of 3D deformations, $p$ is the 3D point to deform, $v$ is the current viewpoint, and $n$ is the number of deformations.

\begin{algorithm}[H]
\caption{Interpolation using the recursive formula \ref{eq:interpolation_point}}
\label{alg:interpolation}
\begin{algorithmic}[1]
\Function{D}{$\{\Phi_j\}$, $p$, $v$, $i$}
    \If{$i = 1$}
        \State \Return $\Phi_1(p)$
    \Else
        \State $p', J \gets $ \Call{D}{$\{\Phi_j\}$, $p$, $v$, $i-1$} \Comment{Get the deformed model}
        \State $\beta \gets B_i(v)$
        \State $p_i, J_i \gets  \Phi_i(p')$
        \State $p'_i \gets \beta \cdot p_i + (1 - \beta) \cdot p'$
        \State $J'_i \gets \beta \cdot J_i \cdot J + (1 - \beta) \cdot J$
        \State \Return $p'_i, J'_i$
    \EndIf
\EndFunction
\end{algorithmic}
\end{algorithm}

%------------------------ 2D to 3D Deformation function -----------------------------%
In the pseudo-code presented in Algorithm~\ref{alg:deformation3d}, $R$ and $T$ represent the transformation matrices that convert world coordinates to camera coordinates. $K$ denotes the intrinsic parameters of the camera, which are used to project points from camera coordinates onto the image plane. All the camera-related matrices are associated with the deformation viewpoint.

\begin{algorithm}[H]
\caption{Apply 2D deformation to 3D}
\label{alg:deformation3d}
\begin{algorithmic}[1]
\Function{$\Phi$}{$p, R, T, K$}
    \State $p_{2d} \gets project(p, R, T, K)$
    \State $p'_{2d} \gets \phi(p_{2d})$ \Comment{new vertex position}
    \State $ J_{2x2} \gets D\phi(p_{2d})$ \Comment{jacobian}
    \State $p', J_{3x3} \gets unproject(p'_{2d}, J_{2x2}, R, T, K)$
    \State \Return $p', J_{3x3}$
\EndFunction
\end{algorithmic}
\end{algorithm}

%------------------------ 2D Deformation function -----------------------------%
In the pseudo-code presented in Algorithm~\ref{alg:deformation2d}, $V$ and $F$ denote the vertices and faces of the 2D mesh, respectively. The set $H$ contains the user-specified vertex indices, known as handles. Each handle $h \in H$ is associated with a corresponding transformation $T_h$, forming the set $\{T_h\}$ of transformations.

\begin{algorithm}[H]
\caption{2D mesh-based deformation}
\label{alg:deformation2d}
\begin{algorithmic}[1]
\Function{$\phi$}{$p_{2d}, V, F, H, \{T_h\}$}
    \For{each triangle $(i, j, k) \in F$}
        \If{$p_{2d}$ is inside triangle $(V_i, V_j, V_k)$}
            \State $t \gets (i, j, k)$ \Comment{Store triangle indices}
            \State \textbf{break}
        \EndIf
    \EndFor

    \State $(\lambda_1, \lambda_2, \lambda_3) \gets \textbf{ComputeBarycentric}(p_{2d}, V_t)$
    
    \State $W \gets \textbf{bbwSolve}(V, F, H)$ \Comment{$|V| \times |H|$ matrix of weights}
    \For{each vertex $v_i \in V$}
        \State $v_i' \gets \sum_{h \in H} W[i, h] \cdot (T_h \cdot v_i)$ \Comment{new vertex positions}
    \EndFor

    \State $p_{2d}' \gets \lambda_1 V'_t[0] + \lambda_2 V'_t[1] + \lambda_3 V'_t[2]$

    \State \Return $p_{2d}'$
\EndFunction
\end{algorithmic}
\end{algorithm}